%% file: main.tex
\documentclass[journal,onecolumn]{IEEEtran}

\usepackage{graphicx}
\usepackage{amssymb}
\usepackage[top=25mm, bottom=25mm, left=25mm, right=25mm]{geometry}
\usepackage{bm}
\usepackage{algorithmic}

\usepackage{caption}
\usepackage{subcaption}            % Required for subfigures.
\usepackage[utf8]{inputenc}
\usepackage[table]{xcolor}
\usepackage{amsmath}
\usepackage{placeins}
\usepackage{mathtools}
\usepackage{dblfloatfix}
\usepackage{float}
\newlength{\topFigureVerticalSpace}
\newlength{\bottomFigureVerticalSpace}
\newcommand{\bfvspace}{\vspace*{-\bottomFigureVerticalSpace}}

\newcommand{\vssubfigure}{\vspace*{2mm}}
\newcommand{\minusvssubfigure}{\vspace*{-2mm}}
\newcolumntype{C}[1]{>{\centering\arraybackslash}p{#1}}

\begin{document}

\title{Computed tomography medical image reconstruction on affordable equipment by using out-of-core techniques}


\author{Mónica Chillarón, Gregorio Quintana-Ortí, Vicente Vidal, and Gumersindo Verdú
\thanks{M. Chillarón and V. Vidal are with the Dpto. de Sistemas Informáticos y Computación, Universitat Politècnica de València, Valencia, 46022 Spain (e-mails: mnichipr@inf.upv.es, vvidal@dsic.upv.es).} 
\thanks{G. Quintana-Ortí is with the Depto. de Ingeniería y Ciencia de Computadores, Universitat Jaume I, Castellón, 12071 Spain (e-mail: gquintan@uji.es).}
\thanks{G. Verdú is with the Dpto. de Ingeniería Química y Nuclear, Universitat Politècnica de València, Valencia, 46022 Spain (e-mail: gverdu@iqn.upv.es).}
\thanks{This research has been supported by “Universitat Politècnica de València”, “Generalitat Valenciana” under PROMETEO/2018/035 and ACIF/2017/075, co-financed by FEDER and FSE funds, and the “Spanish Ministry of Science, Innovation and Universities” under Grant RTI2018-098156-B-C54 co-financed by FEDER funds.}
%\thanks{Manuscript received June -, 2019;}}
\thanks{This work has been submitted to the IEEE for possible publication. Copyright may be transferred without notice, after which this version may no longer be accessible.}}
%\markboth{authors \MakeLowercase{\textit{et al.}}: Computed tomography medical image reconstruction on affordable equipment by using out-of-core techniques}}



% for Transactions on Magnetics papers, we must declare the abstract and
% index terms PRIOR to the title within the \IEEEtitleabstractindextext
% IEEEtran command as these need to go into the title area created by
\IEEEtitleabstractindextext{%
\begin{abstract}
\input{0_Abstract.tex}

\end{abstract}
\begin{IEEEkeywords}
\centering CT, QR factorization, Medical Image, Reconstruction, Out-Of-Core,
Affordable Equipment.
\end{IEEEkeywords}}

\maketitle

\IEEEdisplaynontitleabstractindextext

\input{1_Introduction.tex}

\input{2_Methods.tex}

\input{3_Experimental_study.tex}
\input{4_Conclusions.tex}

%%%% \nocite{golub}

% % use section* for acknowledgment
% \section*{Acknowledgment}


%\newpage
%\section*{References}
\bibliography{main}
\bibliographystyle{IEEEtran}
\ifCLASSOPTIONcaptionsoff
  \newpage
\fi

\end{document}

%% file: 0_Abstract.tex
% =============================================================================
% Abstract
% =============================================================================
%
% He combinado las dos frases siguientes en una sola porque se repetía dos 
% veces la reducción y lo de muchos estudios.
%%%% The reduction of the X-rays dose induced in Computed Tomography (CT) scans 
%%%% is the focus of many studies these days as it is an essential medical test.
%%%% Many techniques have been proposed to
%%%% reconstruct high-quality images using a lesser amount of radiation.
As Computed Tomography (CT) scans are an essential medical test,
many techniques have been proposed to
reconstruct high-quality images
using a smaller amount of radiation.
One approach is to employ algebraic factorization methods to reconstruct
the images, using fewer views than the traditional analytical methods.
However, their main drawback is the high computational cost
and hence the time needed to obtain the images, which is critical in the
daily clinical practice.
For this reason, faster methods for solving this problem are required.
In this paper, we propose a new reconstruction method 
based on the QR factorization
that is very efficient on affordable equipment
(standard multicore processors and standard Solid-State Drives)
by using out-of-core techniques.
Combining both affordable hardware and the new software, 
we can boost the performance of the reconstructions and 
implement a reliable and competitive method 
that reconstructs high-quality CT images quickly.

%% file: 1_Introduction.tex
\vspace{-8pt}
% =============================================================================
\section{Introduction}
%\vspace{-7pt}
\label{sec:introduction}
% =============================================================================

Nowadays, Computed tomography (CT)~\cite{Kak2001} 
is an essential diagnostic medical imaging test in clinical practice.
Although it involves the use of X-rays and hence induces ionizing radiation
in patients, the information provided is critical in many cases.
Therefore, it is extremely important to reduce the radiation dose 
as much as possible, and 
thus prevent patients from absorbing a higher dose than the recommended one.
Otherwise, CTs could be a hazard to them, since it has been proven the X-rays
can be harmful, especially to the most vulnerable patients
\cite{de2009projected,hall2008cancer}.

The traditional CT reconstruction employs analytical methods,
which are based on the 
Filtered Back-Projection (FBP)~\cite{fbp, zhuang2004fan, mori2006combination}.
They require a complete set of projections of an object, 
over 360 degrees of rotation.
They are still the most common methods because of their low computational cost 
and therefore fast reconstruction.
However, reducing the X-ray dose is difficult
when a high-quality image must be obtained.
Several methods~\cite{willemink2013iterative,willemink2013iterative2} 
have been developed that reduce the radiation dose
by minimizing the tube's current or voltage,
and then reconstruct the sinograms with statistical methods
that improve the image quality compared to traditional FBP-based methods.

A common approach to reducing the radiation dose is 
the use of iterative methods, which do not require a complete set of
projections, nor are they restricted in terms of projection angles
\cite{ART, SART, WTD, Flores, flores2014parallel, CHILLARON20171195}.
These type of methods require fewer projections to reconstruct an image.
Nevertheless, they involve a high computational cost, which implies that the
reconstructions are much slower than with previous methods.
Moreover, since these methods are iterative, convergence is not guaranteed, 
nor the number of iterations needed to converge.
Several works~\cite{YanLiu2014, Tang2009,Vandeghinste2013,Zhu2013}
showed the problems of working with few-view limited-angle CT.
The use of few views generates streak artifacts 
that can mask or conceal important parts of the image to be reconstructed, 
which can produce information loss. 
This is potentially harmful since it can lead to wrong diagnosis. 
It also poses a problem for secondary applications of the CT images,
as shown in \cite{Vandeghinste2013}, 
where the reduction of the number of views 
to a minimum number implied an inaccurate segmentation of the blood vessels.
Sechopoulos~\cite{Sechopoulos2013} showed that 
few views led to false positives 
in computer-aided detection for breast mass detection.
Unlike direct methods, 
iterative methods often generate patchy or blocky artifacts 
in the reconstructed images due to overregularization \cite{Tang2009,Qi2015}.

Therefore, direct algebraic methods
such as the QR factorization~\cite{rodriguez2018qr, chillaron2019ct}
have been explored recently.
Although they usually require a greater number of views 
than the iterative ones (as was shown in a previous work~\cite{chillaron2018}),
they are much more accurate when the rank of the weights matrix is complete.
The main drawback of the direct algebraic methods is that
the sparsity of the weights matrix cannot be taken advantage of, 
since the matrix fills in and becomes dense as the process advances.
Therefore, space problems 
because of an insufficient main memory (RAM) can arise.
In this case, it is important to find an efficient approach 
to tackle large problems without having to acquire expensive 
and specialized dedicated equipment, 
which would require a large monetary cost.

In our paper, 
we present a solution to the CT image reconstruction problem
by using the direct solution of linear systems based on the QR factorization.
By employing special high-performance software techniques,
high-quality images are obtained on affordable computers.
%%%% as well as high-performance libraries
%%%% (such as BLAS~\cite{BLAS3}, LAPACK~\cite{LAPACK3} and 
%%%% libFLAME~\cite{libflame_ref,CiSE09}),
%
%%%% The use of these computing techniques
%%%% allowed the use of computers with much smaller main memories, 
%%%% and thus cheaper and affordable.
Without these techniques, the computer required would be very expensive 
(tens of thousands of dollars),
mainly due to the price of the large main memory required to store the data.
With these techniques, 
computers with a price about one order of magnitude smaller can be employed.
A careful application of out-of-core (OOC) techniques
allows to read and write blocks of data from/to the hard drive just when they
are needed for the calculations, 
instead of loading the whole matrices into main memory.
By applying this method, 
as well as some other techniques,
we can solve large-scale problems, 
and therefore a fast reconstruction of CT images with high resolutions 
can be achieved.
Our new implementation is time-efficient and also scalable, 
as can be seen in the results.
In addition, both very high quality and
a reduction in the number of views (and therefore radiation) are achieved,
compared to analytical methods.
In our work,
we have checked that the OOC approach is still valid on much larger matrices
than previous works.
Moreover,
we have assessed the performances on
both traditional hard drives (HDD) and modern Solid-State Drives (SSD).

The document is organized as follows: 
Section~\ref{sec:methods} describes the simulation of our projection data, 
as well as the simulated scanner parameters.
It is also explained how to perform a CT reconstruction using the QR
factorization of the weights matrix.
Besides, the metrics employed to measure the image quality are introduced, 
and the QR factorization and the reconstruction algorithm 
are described in detail.
Section~\ref{sec:experimental_study} assesses our new method in terms of
numerical stability and image quality.
A detailed performance study comparing the different
configurations using two types of hard drives is also included.
To conclude, Section~\ref{sec:conclusions} summarizes and discusses the
advantages of the studied method and proposes future lines of work.

%% file: 2_Methods.tex
% =============================================================================
\section{Methods}
\label{sec:methods}
% =============================================================================
\vspace{-2pt}
% -----------------------------------------------------------------------------
\subsection{CT image reconstruction}
% -----------------------------------------------------------------------------

To reconstruct CT images with an algebraic approach, 
we model the problem as:
\begin{equation}\label{eq:sistema}
\normalsize{
   A X= B + W}
   \vspace{-5pt}
\end{equation}

where 
$A=  \left( a_{i,j}\right) \in \mathbb{R}^{M \times N} $ 
denotes the so-called system matrix, with dimensions $M\times N$.
$A$ is the weights matrix that models the physical scanner, being $a_{i,j}$
the contribution of the $i$-th ray on the $j$-th pixel.
The dimension $M$ is the product of the number of detectors of the CT
scanner multiplied by the number of projections or views taken.
$N$ denotes the resolution of the image ($256 \times 256$ pixels,
$512 \times 512$ pixels, etc.).
$B=(B^j)$ is a matrix of $M\times S$ elements, 
where $S$ is the number of slices to be reconstructed, 
and $B^j$ denotes the column $j$ that will correspond to the $j$-th sinogram.
$X=(X^j)$ is a matrix of dimensions $N\times S$, where $X^j$ is the column
where the reconstructed image corresponding with the $j$-th sinogram 
will be stored.
$W$ is the noise contained in the sinograms, which will not be considered
in this paper.

\begin{table}[t]
\caption{\small{Simulated fan-beam scanner parameters.}}
\label{tab:scanner}
\setlength\extrarowheight{2.5pt}
\centering
\begin{tabular}{|c|c|}
\hline
% \multicolumn{2}{c}{} & \multicolumn{4}{|c|}{LSQR} & \multicolumn{4}{|c|}{LSMR}\\
% \cline{3-10}
% \multicolumn{2}{c}{} & \multicolumn{1}{|c|}{1 core} & \multicolumn{1}{|c|}{2 cores} & \multicolumn{1}{|c|}{4 cores} & \multicolumn{1}{|c|}{8 cores} & \multicolumn{1}{|c|}{1 core} & \multicolumn{1}{|c|}{2 cores} & \multicolumn{1}{|c|}{4 cores} & \multicolumn{1}{|c|}{8 cores} \\
% \cline{3-10}
% \hline
% \multirow{4}{*}{\rotatebox[origin=c]{90}{\parbox[c]{1cm}{\centering nº of slices}}} &
% 1 &0.0601&0.0494&0.0442&0.0416&0.0608&0.0500&0.0446&0.0419 \\
% & 8 &0.4943&0.2606&0.1465&0.0901&0.4967&0.2623&0.1478&0.0916\\
%  & 64 &4.5523&3.0850&1.6356&1.2288&4.6098&2.6468&1.8526&1.4167\\
% & 128 &9.1690&4.9606&2.9690&2.0095&9.2818&5.0265&3.0199&2.0549\\
Source trajectory & 360º circular scan \\  \hline
Scan radius & 75 cm \\  \hline
Source-to-detector distance & 150 cm \\  \hline
X-ray source fan angle & 30º \\  \hline
Number of detectors & 1025 \\  \hline
Pixels of the reconstructed image & $512^2$ \\  \hline
Number of projections & 260 \\
\hline
\end{tabular}
\vspace{-5pt}
\end{table}
\FloatBarrier
The sinograms have been simulated using Joseph method~\cite{Joseph}. 
We modeled a fan-beam scanner, 
using the parameters shown in Table~\ref{tab:scanner}. 
As was mentioned before, the number of projections taken depends 
on the desired reconstruction resolution, 
and it needs to be adjusted so that matrix $A$ has full rank. 
The projections are selected according to Eq.~\ref{eq:angulos}, 
where the symmetry of the projection data is broken
by making an angle shift for every quarter of the circumference 
to improve the rank.
\begin{equation}\label{eq:angulos}
%\footnotesize{
\Theta_{i} = 
\left\{
  \begin{matrix*}[l]
    (360/v)$*$(i$-$1)                        & \textbf{if}\quad 1\leq i\leq (v$/$4)\\
    \Theta_{v/4}$+$0.5$+$(360/v)$*$(i$-$1)   & \textbf{if}\quad (v$/$4)<i\leq(v$/$2)\\
    \Theta_{v/2}$-$0.75$+$(360/v)$*$(i$-$1)  & \textbf{if}\quad (v$/$2)<i\leq(3v$/$4)\\
    \Theta_{3v/4}$-$0.25$+$(360/v)$*$(i$-$1) & \textbf{if}\quad (3v$/$4)<i\leq v
  \end{matrix*}
\right.%}
%\vspace{-3pt}
\end{equation}
\normalsize

To solve the problem in Eq.~\ref{eq:sistema}, 
first 
the QR factorization of $A$ is computed (Eq.~\ref{eq:QR}), 
where $Q$ is orthonormal and $R$ is upper triangular. 
Then, to reconstruct the images, Eq.~\ref{eq:qr1} is employed. 
%\cite{Golub1993}
\begin{equation}\label{eq:QR}
  \normalsize{A= QR}
  \vspace{-4pt}
\end{equation}
\vspace{-7pt}
\begin{equation}\label{eq:qr1}
  \normalsize{
  X = R^{-1}(Q^T B)}
  \vspace{-2pt}
\end{equation}
It is important to note that the QR factorization 
does not need to be computed for every image being generated,
since it does not depend on $B$.
It can be computed just once and, by storing the results,
a lot of computational work can be saved.
\vspace{-10pt}
% -----------------------------------------------------------------------------
\subsection{Image Quality Metrics}
% -----------------------------------------------------------------------------
To measure the quality of the reconstructed images, 
we use two well-established metrics for images: 
PSNR (Peak Signal-To-Noise Ratio) and
SSIM (Structural Similarity Index)~\cite{Hore2010}.
The PSNR metric measures the ratio of the image signal to the noise it
contains.
To calculate it, another metric is used, the so-called Mean Square Error
(MSE), which is calculated according to Eq.~\ref{eq:MSE}, and represents
the mean of the squared error between the reference image $I_{0}$ and the
reconstructed image $I$ ($X$ in our equations).
Once the MSE is calculated, it is used to calculate the PSNR according to
Eq.~\ref{eq:psnr}, in which MAX represents the maximum value that a pixel
can take.
The higher the PSNR value we get, the better the reconstruction obtained.

SSIM measures the internal structures (shapes) of the images
compared with the reference image.
Therefore, it does not focus on the gray levels of the pixels, 
but on the shapes of the reconstructed image 
with respect to the reference image. 
Therefore, it measures what is perceptible to the human eye.
It is applied through windows of fixed size, and the difference between two
windows $x$ and $y$ corresponding to the two images to be compared is
calculated using Eq.~\ref{eq:ssim}.
In this equation, $\mu_{x}$ and $\mu_{y} $ denote the average value of the
window $x$ and $y$, $\sigma^{2}_{x}$ and $\sigma^{2}_{y}$ the variance,
$\sigma_{xy}$ the co-variance between the windows, and c\textsubscript{1}
and c\textsubscript{2} are two stabilizing variables dependent on the
dynamic range of the image.
\begin{equation}\label{eq:MSE}
  \normalsize
    \mbox{MSE} = 
      \frac{1}{MN}\sum_{i=0}^{M-1}\sum_{j=0}^{N-1}(I_0(i,j)-I(i,j))^2
\end{equation}
\begin{equation}\label{eq:psnr}
 \normalsize
  \mbox{PSNR} = 
    10 \log_{10}\frac{\mbox{MAX}(I_0)^2}{\mbox{MSE}}
\end{equation}
\begin{equation}\label{eq:ssim}
 \normalsize
  \mbox{SSIM} = 
  \frac{(2\mu_x\mu_y+c_1)(2\sigma_{x,y}+c_2)}
       {(\mu_x^2+\mu_y^2+c_1)(\sigma_x^2+\sigma_y^2+c_2)}
\end{equation}

% -----------------------------------------------------------------------------
\subsection{Out-of-core computations}
% -----------------------------------------------------------------------------

Some problems require the storage of data 
so large that there are no computers with such a main memory
or, in case they exist, their prices are very high.
Most operating systems provide a virtual memory system 
to store data (and programs) 
that do not fit into the computer's main memory at one time.
However, its performances are not very high
when employed on structured scientific problems.
Hence,
in high-performance scientific computing,
special techniques,
called Out-Of-Core (OOC) or Out-Of-Memory (OOM),
are required to efficiently process data stored in the hard drive.
These techniques keep the data stored in the hard drive, 
read them into memory, and write them into disk whenever is needed.
The aim of these techniques is to minimize the effect of 
the slow speed of the read and write operations from/to disks
in order to render performances as high as possible.

% -----------------------------------
\subsubsection{Traditional approach}
% -----------------------------------

In modern computer architectures 
floating-point operations are much faster than memory accesses.
Therefore, the ratio of flops to memory accesses 
%%%% (\textit{memops})
in computations is very important.
An increased ratio provides much higher performances
since it allows to compute several or even many flops 
per each memory access,
and hence cache memories and other modern features can be fully exploited.
For instance, matrix-matrix operations obtain 
significantly higher performances than matrix-vector operations.
%%%% is so complex that
%%%% operations that are very alike render very different performances.
%%%% In vector-vector operations and matrix-vector operations,
%%%% the ratio of flops to \textit{memops} is usually very low: $\mathcal{O}(1)$
%%%% ($\mathcal{O}(n)$ flops to $\mathcal{O}(n)$ \textit{memops}, and
%%%% $\mathcal{O}(n^2)$ flops to $\mathcal{O}(n^2)$ \textit{memops}, respectively).
%%%% In contrast,
%%%% in matrix-matrix operations
%%%% the ratio of flops to \textit{memops} is much higher: $\mathcal{O}(n)$
%%%% ($\mathcal{O}(n^3)$ flops to $\mathcal{O}(n^2)$ \textit{memops}).
%%%% This increased ratio provides much higher performances
%%%% In fact, a flop in a matrix-vector operation can be 
%%%% about several times slower (between about five and ten times) 
%%%% than a flop in a matrix-matrix operation
%%%% of the same dimensions.

In linear algebra,
unblocked algorithms perform one stage at a time 
(e.g.~one column in column-oriented algorithms).
In contrast,
a blocked algorithm performs several stages 
(e.~g.~several columns in column-oriented algorithms)
of the traditional (unblocked) algorithm at the same time
because this aggregation can take advantage 
of the more efficient matrix-matrix operations.
This number of stages (e.~g.~columns) that are processed at the same time 
is usually called the block size.
%%%% For many years a large and continuous effort has been invested 
%%%% in the development of blocked algorithms
%%%% for performing the most important factorizations and computations 
%%%% in linear algebra.
%%%% In this way, the rewriting of the LINPACK library produced 
%%%% the much more efficient LAPACK library~\cite{WN20,LAPACK3}.

However, 
since most usual algorithms in linear algebra
proceed on triangular matrices,
processing a fixed number of columns (or rows) at the same time
can make the data to be processed very large at the beginning,
and very small at the end, or vice versa.
%
%%%% One good example related to our work is the backward substitution algorithm
%%%% to solve a system of linear equations $R X = B$ 
%%%% with an upper triangular coefficient matrix $R$.
%%%% Let us assume a row-block algorithm is employed.
%%%% In this case, 
%%%% since the first task of the algorithm 
%%%% is to process the right bottom block of $R$
%%%% to compute the bottom block of $X$,
%%%% the amount of data involved is initially very small.
%%%% In contrast, in the final stage, 
%%%% when computing the top block of $X$,
%%%% the amount of data involved is much larger
%%%% since the whole $X$ and all the first block row of $R$ are accessed.
%
This can make performances not to be optimal because 
main memory could be underused in some stages and 
because of the large variation in the data being transferred.
This variation of the transfer size 
can be a problem when the data are stored in disk
since this kind of devices are more sensitive to transfer sizes.
%
%%%% The size of the block size must be selected 
%%%% so that at least two block columns can fit in memory.
%%%% A large fixed block size (the usual approach) makes this very efficient
%%%% at the beginning (when all the memory is employed),
%%%% but very inefficient in the last stages 
%%%% when the amount of data are not so large (or vice versa).

There are usually two common types of algorithms:
Right-looking algorithms update the rest of the matrix (right part)
after the processing of the current (block) column or row,
thus requiring $\mathcal{O}(n^3)$ writes.
In contrast,
left-looking algorithms update the current (block) column or row,
with the data previously processed (left part),
thus requiring $\mathcal{O}(n^2)$ writes.
Since the cost of a write operation in hard drives 
is usually higher than the cost of a read operation,
left-looking algorithms are usually preferred 
when working on data stored in disk.
Great efforts have been made to efficiently solve problems
from linear algebra
whose data do not fit in RAM and must be stored 
in disk~\cite{SOLAR,ooc:scalapack,OOC:TR,Gunter:2005:POC,OOCLU:Para04,OOC:PDSECA-01}.

% -----------------------------------
\subsubsection{Algorithms-by-blocks}
% -----------------------------------

Like blocked algorithms,
algorithms-by-blocks also perform several stages of the 
traditional (unblocked) algorithm at the same time
in order to take advantage of the higher speeds of matrix-matrix operations.
Unlike blocked algorithms,
Algorithms-by-blocks achieve matrix-matrix operations 
by raising the granularity of the data.
First, the traditional (unblocked) algorithm must be reformulated 
to perform operations that process only scalar elements.
Then, the scalar elements are raised 
to being square blocks of dimension $b \times b$,
and the operations processing them are accordingly raised too
so that they correctly process these square blocks.
Therefore, 
in the end the whole computation to be performed is divided
into many tasks, each one processing a few square blocks
(between one and four, but more usually two or three).

One of the main benefits of this approach is that
all blocks are always of the same size 
(except maybe for the final right and bottom blocks).
%%%% if the block size is not a multiple of the matrix dimensions).
This brings in the benefit of making 
the majority of the transfers of the same size.
Thus, by tuning the block size for a given machine,
all the data transfers will be very efficient,
regardless of the stage of the algorithm (first stages or last stages).

Quintana-Ort\'{\i} \textit{et al.}~\cite{OOC:TOMS,ooc:supercomputing}
developed a runtime that can process algorithm-by-blocks 
very efficiently by applying two techniques:
The use of a cache of blocks stored in memory 
to reuse information,
and the overlapping of computation and communications
to reduce the cost of the latter.

% -------------------------------
\subsubsection{QR factorization}
% -------------------------------

The algorithm-by-blocks for efficiently computing the QR factorization 
was described in 2009~\cite{ooc:qr}.
This approach employed the methods and runtime described by
Quintana-Ort\'{\i} \textit{et al.}~\cite{OOC:TOMS,ooc:supercomputing}.
However,
these works assessed smaller matrices, 
they did not test modern fast Solid-State Drives (SSD),
and they only assessed the QR factorization.
In our current work we have checked 
that this approach is still valid on much larger matrices,
we have compared the performances of this approach on 
both traditional hard drives and modern SSDs,
and we have implemented and assessed the application of orthogonal 
transformations previously computed and 
the resolution of triangular linear systems
(problems not included in these previous works).

%%%% We are describing the out-of-core algorithm-by-blocks for computing
%%%% the QR factorization
%%%% because the system solving will apply the same techniques,
%%%% and even use some building blocks.

Figure~\ref{fig:qr_ab}
illustrates the process performed by a left-looking algorithm-by-blocks
for computing the QR factorization of a $9 \times 9$ matrix
with block size 3.
The `$\bullet$' symbol represents a non-modified element by the current task,
the `$\ast$' symbol represents a modified element by the current task, and
the `$\circ$' symbol represents a nullified element
(either by the current task or by a previous task).
The nullified elements are shown because they store information 
about the Householder transformations that will be later used to apply them.
The continuous lines surround the blocks involved in the current task.
To reduce the size of this graphic,
it only shows the factorization of the first and second block columns.
In the processing of the first column, as there are no previous columns,
the work to do is just to nullify all the elements below the main diagonal.
This process is performed with three tasks (tasks 1, 2, and 3).
The first task nullifies elements below the diagonal in $A_{00}$.
The second and third tasks nullify elements in $A_{10}$ and $A_{20}$, 
respectively.
To nullify those two blocks,
these two tasks must also update the $A_{00}$ block.
In the processing of the second column, 
the first work to do is to apply previous transformations 
to the current block column (tasks 4, 5, and 6).
Then, the elements below the diagonal in blocks $A_{11}$ and $A_{21}$ 
must be nullified (tasks 7 and 8).

\input{fig_tasks_qr_ab.tex}

Table~\ref{fig:tasks_qr_ab} illustrates 
all the tasks generated and executed 
by the algorithm-by-blocks for computing the QR factorization
for the previous case (and also for the general cases $m = n = 3 b$, 
where $b$ is the block size).
As can be seen, in this case the full factorization comprises 14 tasks.
The effect of the first eight tasks is shown in Figure~\ref{fig:qr_ab}.
The remaining tasks (not shown in the graphic)
proceed in an analogous way on the third block column:
First, 
the transformations 
obtained when annihilating the elements below the diagonal 
in the first block column
are applied to the third block column.
Second,
the transformations 
obtained when annihilating the elements below the diagonal 
in the second block column
are applied to the third block column.
Finally, 
the elements below the diagonal in the third block column are annihilated.

\input{fig_qr_ab.tex}
%\par\bigskip 
 \noindent
% \nopagebreak[4]

The QR factorization of a matrix of any dimension only requires 
the following four generic tasks:

\begin{itemize}

\item
\textit{Compute\_dense\_QR( A, S )}:
This task nullifies all the elements below the diagonal 
of input/output block $A$.
The output is two-fold:
The first is the updated matrix $A$, and 
the second is the $S$ factor.
The upper triangular part of $A$ contains the updated $R$ triangular factor.
The strictly lower triangular part of $A$ contains the Householder reflectors
generated in this QR factorization.
Matrix $S$ contains the $S$ factors,
also required to apply the transformations obtained in this task.

\item
\textit{Apply\_left\_Qt\_of\_dense\_QR( Y, S, C )}:
The input data of this task are 
matrices $Y$ (the Householder reflectors)
and $S$ (the $S$ factors),
the output of the previous task.
Given these two input matrices $Y$ and $S$,
this task applies those transformations to input/output block $C$.

\item
\textit{Compute\_TD\_QR( T, D, S )}:
The input data of this task are matrices $T$ and $D$
(triangular and dense, respectively, and hence the acronym TD).
This task nullifies all the elements in block $D$ 
and accordingly updates block $T$.
The output is three-fold:
The first output is matrix $T$ (containing the updated triangular factor),
the second output is matrix $D$ (containing the Householder reflectors), and
the third output is matrix $S$ (containing the $S$ factors).

\item
\textit{Apply\_left\_Qt\_of\_TD\_QR( D, S, F, G )}:
The input data of this task are 
the input matrix $D$ (the Householder reflectors) and $S$ (the $S$ factors).
Both of them are the output of the previous task,
i.e.~the computation of the QR factorization of a triangular-dense factor.
This task correspondingly updates input/output matrices $F$ and $G$ 
with those transformations.

\end{itemize}

These four generic tasks will be employed
when computing the QR factorization ($A = QR$) and 
when computing the solution of the linear system $X = R^{-1} (Q^T B)$.

% -----------------------------
\subsubsection{System solving}
% -----------------------------

When a linear system of equations $A X = B$ must be solved
by using the QR factorization,
the first stage is obviously to compute the factorization $A = QR$.
The second stage is the following computation: 
$X = R^{-1} (Q^T B)$,
where $Q^T$ is the transpose of $Q$.

The first sub-step of the second stage 
($X = R^{-1} (Q^T B)$)
is to compute the product $Q^T B$.
As usual in linear algebra, 
matrix $Q$ (or its transpose) is not explicitly built
because of the large cost (in both space and time) of the building operation 
and the even larger computational cost of the following matrix-matrix multiply.
Instead, the transpose of matrix $Q$ will be implicitly applied 
by using the Householder reflectors and the $S$ factors 
previously obtained in the QR factorization.
\FloatBarrier

The second sub-step of the second stage 
($X = R^{-1} (Q^T B)$)
is to multiply
the inverse of $R$ and the result of the previous sub-step ($Q^T B$).
As usual in linear algebra, 
to reduce the computational cost,
the inverse of $R$ is not explicitly computed, and 
instead a linear backward substitution is applied.
A block row algorithm for the backward substitution
has been employed in order to 
both increase the locality 
and minimize the number of blocks being written 
(if a cache of blocks is employed).

Table~\ref{fig:tasks_solve_ab}
illustrates all the tasks generated and executed by 
the algorithm-by-blocks for computing 
$X = R^{-1} (Q^T B)$
when the QR factorization has been previously computed
for the case $m = n = 3 b$,
where $b$ is the block size.

\input{fig_tasks_solve_ab.tex}

\FloatBarrier

%% file: fig_tasks_qr_ab.tex
\begin{table}[!t]
\caption{\small{List of tasks generated by the Algorithm-by-blocks 
for computing the QR factorization when $m = n = 3 b$,
where $b$ is the block size.}}
\label{fig:tasks_qr_ab}
\centering
\setlength\tabcolsep{2pt}
\begin{tabular}{|l|c|c|}
  \hline
  \multicolumn{1}{|c|}{Operation} & \multicolumn{2}{c|}{Operands} \\ \cline{2-3}
  & Out & In \\ \hline
  Comp\_dense\_QR & 
      $A_{00}$ $S_{00}$ & 
      $A_{00}$ \\ \hline
  Comp\_TD\_QR & 
      $A_{00}$ $A_{10}$ $S_{10}$ &
      $A_{00}$ $A_{10}$ \\ \hline
  Comp\_TD\_QR & 
      $A_{00}$ $A_{20}$ $S_{20}$ &
      $A_{00}$ $A_{20}$ \\ \hline
  Apply\_left\_Qt\_of\_dense\_QR &
      $A_{01}$ &
      $A_{00}$ $S_{00}$ $A_{01}$ \\ \hline
  Apply\_left\_Qt\_of\_TD\_QR&
      $A_{01}$ $A_{11}$ &
      $A_{10}$ $S_{10}$ $A_{01}$ $A_{11}$ \\ \hline
  Apply\_left\_Qt\_of\_TD\_QR &
      $A_{01}$ $A_{21}$ &
      $A_{20}$ $S_{20}$ $A_{01}$ $A_{21}$ \\ \hline
  Comp\_dense\_QR &
      $A_{11}$ $S_{11}$ & 
      $A_{11}$ \\ \hline
  Comp\_TD\_QR &
      $A_{11}$ $A_{21}$ $S_{21}$ & 
      $A_{11}$ $A_{21}$ \\ \hline
  Apply\_left\_Qt\_of\_dense\_QR &
      $A_{02}$ & 
      $A_{00}$ $S_{00}$ $A_{02}$ \\ \hline
  Apply\_left\_Qt\_of\_TD\_QR &
      $A_{02}$ $A_{12}$ & 
      $A_{10}$ $S_{10}$ $A_{02}$ $A_{12}$ \\ \hline
  Apply\_left\_Qt\_of\_TD\_QR &
      $A_{02}$ $A_{22}$ & 
      $A_{20}$ $S_{20}$ $A_{02}$ $A_{22}$ \\ \hline
  Apply\_left\_Qt\_of\_dense\_QR &
      $A_{12}$ & 
      $A_{11}$ $S_{11}$ $A_{12}$ \\ \hline
  Apply\_left\_Qt\_of\_TD\_QR &
      $A_{12}$ $A_{22}$ & 
      $A_{21}$ $S_{21}$ $A_{12}$ $A_{22}$ \\ \hline
  Comp\_dense\_QR &
      $A_{22}$ $S_{22}$ & 
      $A_{22}$ \\ \hline
\end{tabular}
\vspace{-5pt}
\end{table}

%% file: fig_qr_ab.tex
%
% Figure with tasks of QR_AB.
%

% Usual element:  Usual bullet. 
\newcommand{\mybl}{\bullet}
% Usual element:  Usual bullet forced not to have surrounding lines.
\newcommand{\mybn}{\multicolumn{1}{c}{\bullet}} 
% Modified element:  Usual asterisk. 
\newcommand{\mysl}{\ast}
% Modified element:  Usual asterisk forced not to have surrounding lines.
\newcommand{\mysn}{\multicolumn{1}{c}{\ast}}
% Nullified element:  Usual circ. 
\newcommand{\mypl}{\circ}
% Nullified element:  Usual circ forced not to have surrounding lines.
\newcommand{\mypn}{\multicolumn{1}{c}{\circ}}

% Use arabic numbers for the subfigures.
\renewcommand{\thesubfigure}{\arabic{subfigure}}

%\onecolumn
\begin{figure*}[]
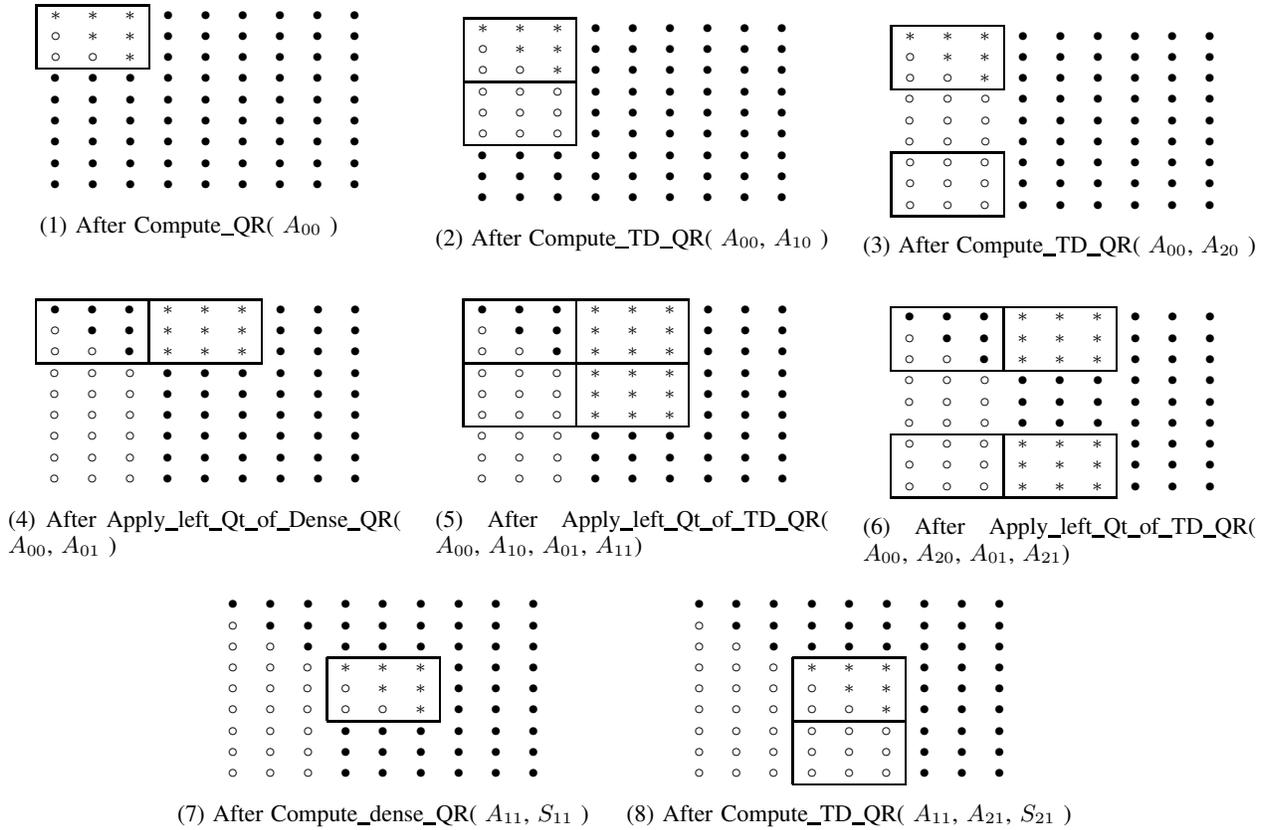

\vspace{-10pt}
\footnotesize
\centering

%%%% \begin{subfigure}{.30\linewidth}
%%%% \[
%%%% \begin{array}{ccccccccc}
%%%%   \mybl & \mybl & \mybl & \mybl & \mybl & \mybl & \mybl & \mybl & \mybl \\
%%%%   \mybl & \mybl & \mybl & \mybl & \mybl & \mybl & \mybl & \mybl & \mybl \\
%%%%   \mybl & \mybl & \mybl & \mybl & \mybl & \mybl & \mybl & \mybl & \mybl \\
%%%%   \mybl & \mybl & \mybl & \mybl & \mybl & \mybl & \mybl & \mybl & \mybl \\
%%%%   \mybl & \mybl & \mybl & \mybl & \mybl & \mybl & \mybl & \mybl & \mybl \\
%%%%   \mybl & \mybl & \mybl & \mybl & \mybl & \mybl & \mybl & \mybl & \mybl \\
%%%%   \mybl & \mybl & \mybl & \mybl & \mybl & \mybl & \mybl & \mybl & \mybl \\
%%%%   \mybl & \mybl & \mybl & \mybl & \mybl & \mybl & \mybl & \mybl & \mybl \\
%%%%   \mybl & \mybl & \mybl & \mybl & \mybl & \mybl & \mybl & \mybl & \mybl \\
%%%% \end{array}
%%%% \]
%%%% \minusvssubfigure
%%%% \caption{Initial matrix \newline}
%%%% \vssubfigure
%%%% \end{subfigure}
%%%% %
%%%% %
%%%% \hspace*{0.3cm}
%%%% %
%%%% %
\scriptsize{
\begin{subfigure}{.315\linewidth}
\centering
\[
\begin{array}{|ccc|cccccc} 
\cline{1-3}
  \mysl & \mysl & \mysl & \mybl & \mybl & \mybl & \mybl & \mybl & \mybl \\
  \mypl & \mysl & \mysl & \mybl & \mybl & \mybl & \mybl & \mybl & \mybl \\
  \mypl & \mypl & \mysl & \mybl & \mybl & \mybl & \mybl & \mybl & \mybl \\ 
\cline{1-3}
  \mybn & \mybl & \mybn & \mybl & \mybl & \mybl & \mybl & \mybl & \mybl \\
  \mybn & \mybl & \mybn & \mybl & \mybl & \mybl & \mybl & \mybl & \mybl \\
  \mybn & \mybl & \mybn & \mybl & \mybl & \mybl & \mybl & \mybl & \mybl \\
  \mybn & \mybl & \mybn & \mybl & \mybl & \mybl & \mybl & \mybl & \mybl \\
  \mybn & \mybl & \mybn & \mybl & \mybl & \mybl & \mybl & \mybl & \mybl \\
  \mybn & \mybl & \mybn & \mybl & \mybl & \mybl & \mybl & \mybl & \mybl \\
\end{array}
\]
\minusvssubfigure
\caption{After Compute\_QR( $A_{00}$ ) \newline}
\vssubfigure
\end{subfigure}
\hspace*{0.3cm}
\begin{subfigure}{.315\linewidth}
\centering
\[
\begin{array}{|ccc|cccccc} 
\cline{1-3}
  \mysl & \mysl & \mysl & \mybl & \mybl & \mybl & \mybl & \mybl & \mybl \\
  \mypl & \mysl & \mysl & \mybl & \mybl & \mybl & \mybl & \mybl & \mybl \\
  \mypl & \mypl & \mysl & \mybl & \mybl & \mybl & \mybl & \mybl & \mybl \\ 
\cline{1-3}
  \mypl & \mypl & \mypl & \mybl & \mybl & \mybn & \mybl & \mybl & \mybn \\
  \mypl & \mypl & \mypl & \mybl & \mybl & \mybn & \mybl & \mybl & \mybn \\
  \mypl & \mypl & \mypl & \mybl & \mybl & \mybn & \mybl & \mybl & \mybn \\
\cline{1-3}
  \mybn & \mybl & \mybn & \mybl & \mybl & \mybn & \mybl & \mybl & \mybn \\
  \mybn & \mybl & \mybn & \mybl & \mybl & \mybn & \mybl & \mybl & \mybn \\
  \mybn & \mybl & \mybn & \mybl & \mybl & \mybn & \mybl & \mybl & \mybn \\
\end{array}
\]
\minusvssubfigure
\caption{After Compute\_TD\_\-QR( $A_{00}$, $A_{10}$ )}
\vssubfigure
\end{subfigure}
\hspace*{0.3cm}
\begin{subfigure}{.315\linewidth}
\centering
\[
\begin{array}{|ccc|cccccc} 
\cline{1-3}
  \mysl & \mysl & \mysl & \mybl & \mybl & \mybl & \mybl & \mybl & \mybl \\
  \mypl & \mysl & \mysl & \mybl & \mybl & \mybl & \mybl & \mybl & \mybl \\
  \mypl & \mypl & \mysl & \mybl & \mybl & \mybl & \mybl & \mybl & \mybl \\ 
\cline{1-3}
  \mypn & \mypl & \mypn & \mybl & \mybl & \mybn & \mybl & \mybl & \mybn \\
  \mypn & \mypl & \mypn & \mybl & \mybl & \mybn & \mybl & \mybl & \mybn \\
  \mypn & \mypl & \mypn & \mybl & \mybl & \mybn & \mybl & \mybl & \mybn \\
\cline{1-3}
  \mypl & \mypl & \mypl & \mybl & \mybl & \mybn & \mybl & \mybl & \mybn \\
  \mypl & \mypl & \mypl & \mybl & \mybl & \mybn & \mybl & \mybl & \mybn \\
  \mypl & \mypl & \mypl & \mybl & \mybl & \mybn & \mybl & \mybl & \mybn \\
\cline{1-3}
\end{array}
\]
\minusvssubfigure
\caption{After Compute\_TD\_\-QR( $A_{00}$, $A_{20}$ )}
\end{subfigure}
%
%
% \hspace*{0.3cm}
%
%
\begin{subfigure}{.315\linewidth}
\centering
\[
\begin{array}{|ccc|ccc|ccc} 
\cline{1-6}
  \mybl & \mybl & \mybl & \mysl & \mysl & \mysl & \mybl & \mybl & \mybl \\
  \mypl & \mybl & \mybl & \mysl & \mysl & \mysl & \mybl & \mybl & \mybl \\
  \mypl & \mypl & \mybl & \mysl & \mysl & \mysl & \mybl & \mybl & \mybl \\ 
\cline{1-6}
  \mypn & \mypl & \mypn & \mybl & \mybl & \mybn & \mybl & \mybl & \mybl \\
  \mypn & \mypl & \mypn & \mybl & \mybl & \mybn & \mybl & \mybl & \mybl \\
  \mypn & \mypl & \mypn & \mybl & \mybl & \mybn & \mybl & \mybl & \mybl \\
  \mypn & \mypl & \mypn & \mybl & \mybl & \mybn & \mybl & \mybl & \mybl \\
  \mypn & \mypl & \mypn & \mybl & \mybl & \mybn & \mybl & \mybl & \mybl \\
  \mypn & \mypl & \mypn & \mybl & \mybl & \mybn & \mybl & \mybl & \mybl \\
\end{array}
\]
\minusvssubfigure
\caption{After Apply\_left\_Qt\_of\_\-Den\-se\_QR( $A_{00}$, $A_{01}$ )}
\vssubfigure
\end{subfigure}
\hspace*{0.3cm}
\begin{subfigure}{.315\linewidth}
\centering
\[
\begin{array}{|ccc|ccc|ccc} 
\cline{1-6}
  \mybl & \mybl & \mybl & \mysl & \mysl & \mysl & \mybl & \mybl & \mybl \\
  \mypl & \mybl & \mybl & \mysl & \mysl & \mysl & \mybl & \mybl & \mybl \\
  \mypl & \mypl & \mybl & \mysl & \mysl & \mysl & \mybl & \mybl & \mybl \\ 
\cline{1-6}
  \mypl & \mypl & \mypl & \mysl & \mysl & \mysl & \mybl & \mybl & \mybn \\
  \mypl & \mypl & \mypl & \mysl & \mysl & \mysl & \mybl & \mybl & \mybn \\
  \mypl & \mypl & \mypl & \mysl & \mysl & \mysl & \mybl & \mybl & \mybn \\
\cline{1-6}
  \mypn & \mypl & \mypn & \mybl & \mybl & \mybn & \mybl & \mybl & \mybn \\
  \mypn & \mypl & \mypn & \mybl & \mybl & \mybn & \mybl & \mybl & \mybn \\
  \mypn & \mypl & \mypn & \mybl & \mybl & \mybn & \mybl & \mybl & \mybn \\
\end{array}
\]
\minusvssubfigure
\caption{After Apply\_left\_Qt\_of\_\-TD\-\_QR( $A_{00}$, $A_{10}$, $A_{01}$, $A_{11}$)}
\vssubfigure
\end{subfigure}
\hspace*{0.3cm}
\begin{subfigure}{.315\linewidth}
\centering
\[
\begin{array}{|ccc|ccc|ccc} 
\cline{1-6}
  \mybl & \mybl & \mybl & \mysl & \mysl & \mysl & \mybl & \mybl & \mybl \\
  \mypl & \mybl & \mybl & \mysl & \mysl & \mysl & \mybl & \mybl & \mybl \\
  \mypl & \mypl & \mybl & \mysl & \mysl & \mysl & \mybl & \mybl & \mybl \\ 
\cline{1-6}
  \mypn & \mypl & \mypn & \mybl & \mybl & \mybn & \mybl & \mybl & \mybn \\
  \mypn & \mypl & \mypn & \mybl & \mybl & \mybn & \mybl & \mybl & \mybn \\
  \mypn & \mypl & \mypn & \mybl & \mybl & \mybn & \mybl & \mybl & \mybn \\
\cline{1-6}
  \mypl & \mypl & \mypl & \mysl & \mysl & \mysl & \mybl & \mybl & \mybn \\
  \mypl & \mypl & \mypl & \mysl & \mysl & \mysl & \mybl & \mybl & \mybn \\
  \mypl & \mypl & \mypl & \mysl & \mysl & \mysl & \mybl & \mybl & \mybn \\
\cline{1-6}
\end{array}
\]
\minusvssubfigure
\caption{After Apply\_left\_Qt\_of\_\-TD\_QR( $A_{00}$, $A_{20}$, $A_{01}$, $A_{21}$)}
\end{subfigure}
%
%
% \hspace*{0.3cm}
%
%
\begin{subfigure}{.33\linewidth}
\centering
\[
\begin{array}{ccc|ccc|ccc} 
  \mybn & \mybn & \mybn & \mybn & \mybn & \mybn & \mybl & \mybl & \mybl \\
  \mypn & \mybn & \mybn & \mybn & \mybn & \mybn & \mybl & \mybl & \mybl \\
  \mypn & \mypn & \mybn & \mybn & \mybn & \mybn & \mybl & \mybl & \mybl \\ 
\cline{4-6} 
  \mypn & \mypl & \mypl & \mysl & \mysl & \mysl & \mybl & \mybl & \mybn \\
  \mypn & \mypl & \mypl & \mypl & \mysl & \mysl & \mybl & \mybl & \mybn \\
  \mypn & \mypl & \mypl & \mypl & \mypl & \mysl & \mybl & \mybl & \mybn \\
\cline{4-6} 
  \mypn & \mypl & \mypn & \mybn & \mybn & \mybn & \mybl & \mybl & \mybn \\
  \mypn & \mypl & \mypn & \mybn & \mybn & \mybn & \mybl & \mybl & \mybn \\
  \mypn & \mypl & \mypn & \mybn & \mybn & \mybn & \mybl & \mybl & \mybn \\
\end{array}
\]
\minusvssubfigure
\caption{After Compute\_den\-se\_\-QR( $A_{11}$, $S_{11}$ )}
\vssubfigure
\end{subfigure}
\hspace*{0.3cm}
\begin{subfigure}{.36\linewidth}
\centering
\[
\begin{array}{ccc|ccc|ccc} 
  \mybl & \mybl & \mybn & \mybn & \mybl & \mybn & \mybl & \mybl & \mybl \\
  \mypl & \mybl & \mybn & \mybn & \mybl & \mybn & \mybl & \mybl & \mybl \\
  \mypl & \mypl & \mybn & \mybn & \mybl & \mybn & \mybl & \mybl & \mybl \\ 
\cline{4-6}
  \mypl & \mypl & \mypl & \mysl & \mysl & \mysl & \mybl & \mybl & \mybl \\
  \mypl & \mypl & \mypl & \mypl & \mysl & \mysl & \mybl & \mybl & \mybl \\
  \mypl & \mypl & \mypl & \mypl & \mypl & \mysl & \mybl & \mybl & \mybl \\
\cline{4-6}
  \mypn & \mypl & \mypl & \mypl & \mypl & \mypl & \mybl & \mybl & \mybn \\
  \mypn & \mypl & \mypl & \mypl & \mypl & \mypl & \mybl & \mybl & \mybn \\
  \mypn & \mypl & \mypl & \mypl & \mypl & \mypl & \mybl & \mybl & \mybn \\
\cline{4-6}
\end{array}
\]
\minusvssubfigure
\caption{After Compute\_TD\_QR( $A_{11}$, $A_{21}$, $S_{21}$ )}
\vssubfigure
\end{subfigure}
}
\vspace{-5pt}
\caption{\small{An illustration of the first tasks performed by an algorithm-by-blocks for computing the QR factorization. 
The `$\bullet$' symbol represents a non-modified element by the current task,
`$\ast$' represents a modified element by the current task, and
`$\circ$' represents a nullified element 
(by the current task or by a previous task).
The continuous lines surround the blocks involved in the current task.}}
\label{fig:qr_ab}
%\vspace{-5pt}
\end{figure*}
%\twocolumn

%% file: fig_tasks_solve_ab.tex
%\begin{figure}

\begin{table}[t!]
\centering
\setlength\tabcolsep{2pt}
 \caption{\small{List of tasks generated by the Algorithm-by-blocks 
for solving a linear system using a previously computed QR factorization 
when $m = n = 3 b$, where $b$ is the block size}}
    \label{fig:tasks_solve_ab}
\begin{tabular}{|l|c|c|}
  \hline
  \multicolumn{1}{|c|}{Operation} & \multicolumn{2}{c|}{Operands} \\ \cline{2-3}
  & Out & In \\ \hline
  Apply\_left\_Qt\_of\_dense\_QR &
      $B_{00}$& 
      $A_{00}$ $S_{00}$ $B_{00}$ \\ \hline
  Apply\_left\_Qt\_of\_TD\_QR &
      $B_{00}$ $B_{10}$& 
      $A_{10}$ $S_{10}$ $B_{00}$ $B_{10}$ \\ \hline
  Apply\_left\_Qt\_of\_TD\_QR &
      $B_{00}$ $B_{20}$& 
      $A_{20}$ $S_{20}$ $B_{00}$ $B_{20}$ \\ \hline
  Apply\_left\_Qt\_of\_dense\_QR &
      $B_{10}$ & 
      $A_{11}$ $S_{11}$ $B_{10}$ \\ \hline
  Apply\_left\_Qt\_of\_TD\_QR &
      $B_{10}$ $B_{20}$& 
      $A_{21}$ $S_{21}$ $B_{10}$ $B_{20}$ \\ \hline
  Apply\_left\_Qt\_of\_dense\_QR &
      $B_{20}$ & 
      $A_{22}$ $S_{22}$ $B_{20}$ \\ \hline
  Trsm\_lunn ($B = \textnormal{upper}(A)^{-1} B$) &
      $B_{20}$ & 
      $A_{22}$ $B_{20}$ \\ \hline
  Gemm\_nn\_mo ($C = - A B + C$) &
      $B_{10}$ & 
      $B_{10}$ $A_{12}$ $B_{20}$ \\ \hline
  Trsm\_lunn ($B = \textnormal{upper}(A)^{-1} B$) &
      $B_{10}$ & 
      $A_{11}$ $B_{10}$ \\ \hline
  Gemm\_nn\_mo ($C = - A B + C$) &
      $B_{00}$ & 
      $B_{00}$ $A_{01}$ $B_{10}$ \\ \hline
  Gemm\_nn\_mo ($C = - A B + C$) &
      $B_{00}$ & 
      $B_{00}$ $A_{02}$ $B_{20}$ \\ \hline
  Trsm\_lunn ($B = \textnormal{upper}(A)^{-1} B $) &
      $B_{00}$ & 
      $A_{00}$ $B_{00}$ \\ \hline
\end{tabular}
\vspace{-5pt}
\end{table}

%% file: 3_Experimental_study.tex
% =============================================================================
% \vspace{-8pt}
\section{Experimental study}
\label{sec:experimental_study}
% =============================================================================
%\vspace{-2pt}
In this section, 
the precision and the speed of our new implementations are assessed.
The first subsection describes the precision study,
whereas the second subsection describes the performance study.
In all the experiments 
we used double-precision arithmetic with double-precision real matrices.
% \begin{figure}[b!]
% %\tfvspace
% \vspace{-18pt}
% \begin{center}
% \includegraphics[trim={1.4cm 0 1cm 0},clip,width=0.48\textwidth]{plots/residual.eps} \\
% \end{center}
% %\bfvspace
% \vspace{-10pt}
% \caption{\small{Evolution of the relative residual.}}
% \label{fig:residual}
% \vspace{-6pt}
% \end{figure}
%\FloatBarrier
  \vspace{-10pt}

% -----------------------------------------------------------------------------
\subsection{Precision and image quality study}
\label{subsec:precision}
% -----------------------------------------------------------------------------

In a preliminary test to check the validity of this method, 
the Forbild Head Phantom \cite{forbild} for different resolutions (from $64^2$ to $512^2$)
was projected and reconstructed.
Table~\ref{tab:residual} shows the relative residual 
$r=||AX-B||_F / ||A||_F$ for those resolutions.
As can be seen, the data shows that the method is numerically stable and 
the solution obtained is very accurate even on the highest resolution.
Although the residual grows with the resolution, it is still low, 
so higher image resolutions could be reached if needed.

\begin{figure}[h]
\begin{subfigure}{0.24\textwidth}
\centering
  \includegraphics[trim={14cm 4.5cm 14cm 1.5cm},clip,width=3.5cm]{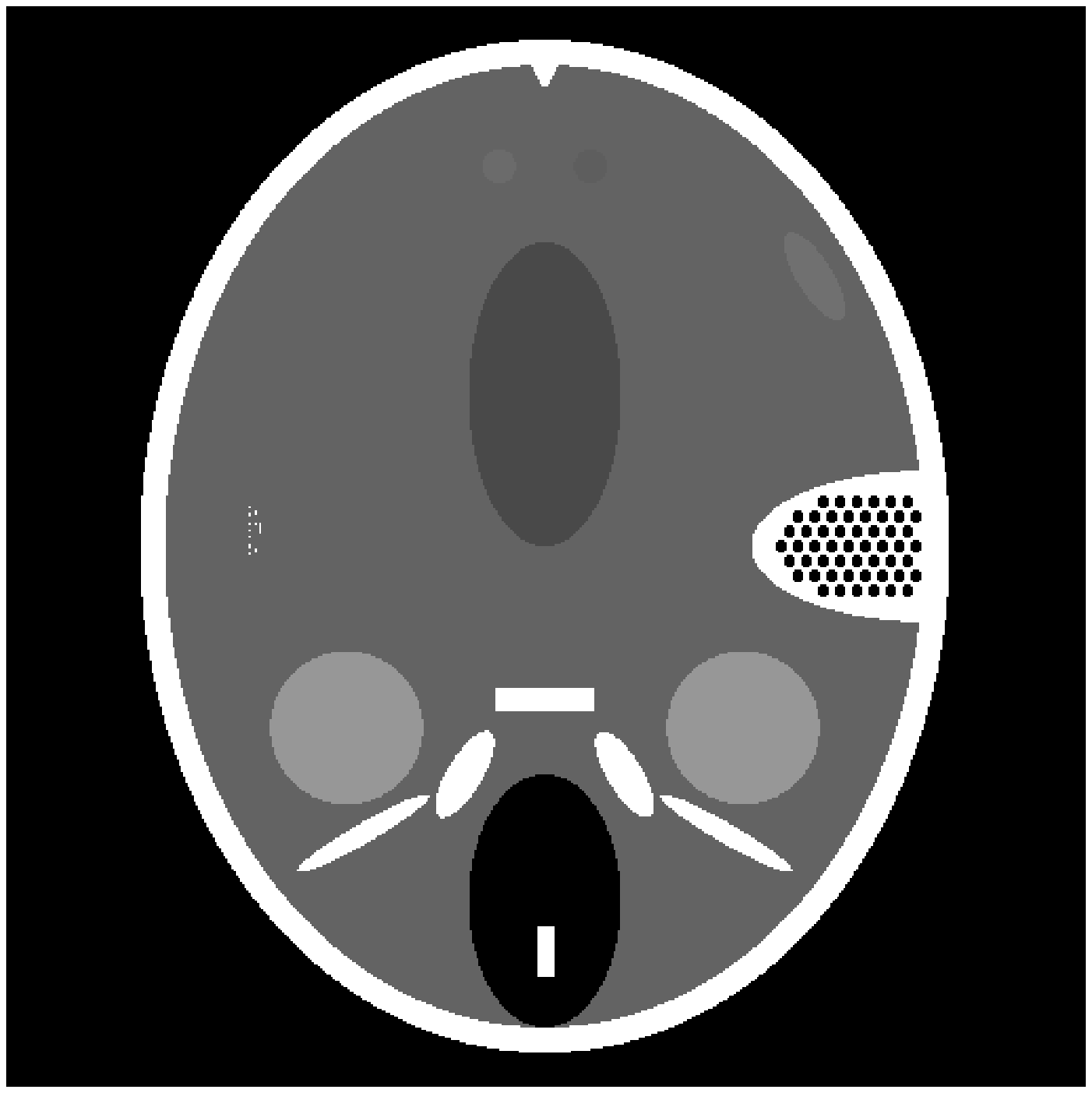}
\tiny\caption{Phantom reference\label{fig:ref}}

\end{subfigure}
\begin{subfigure}{.24\textwidth}
\centering
  \includegraphics[trim={14cm 4.5cm 14cm 1.5cm},clip,width=3.5cm]{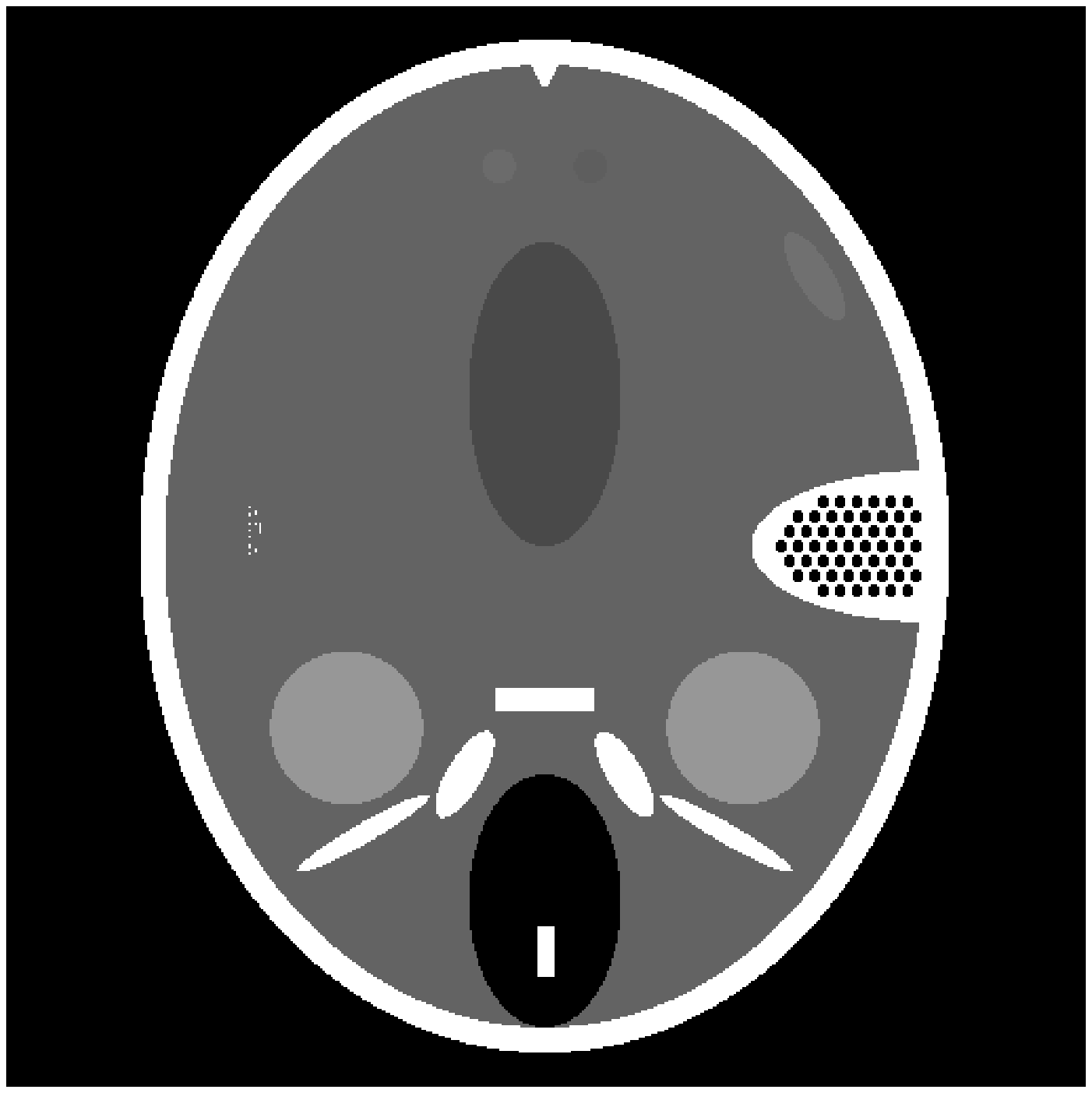}
\caption{QR reconstruction\label{fig:rec}}

\end{subfigure}
\begin{subfigure}{0.24\textwidth}
\centering
  \includegraphics[trim={3.5cm 3.5cm 3.5cm 3.5cm},clip,width=4cm]{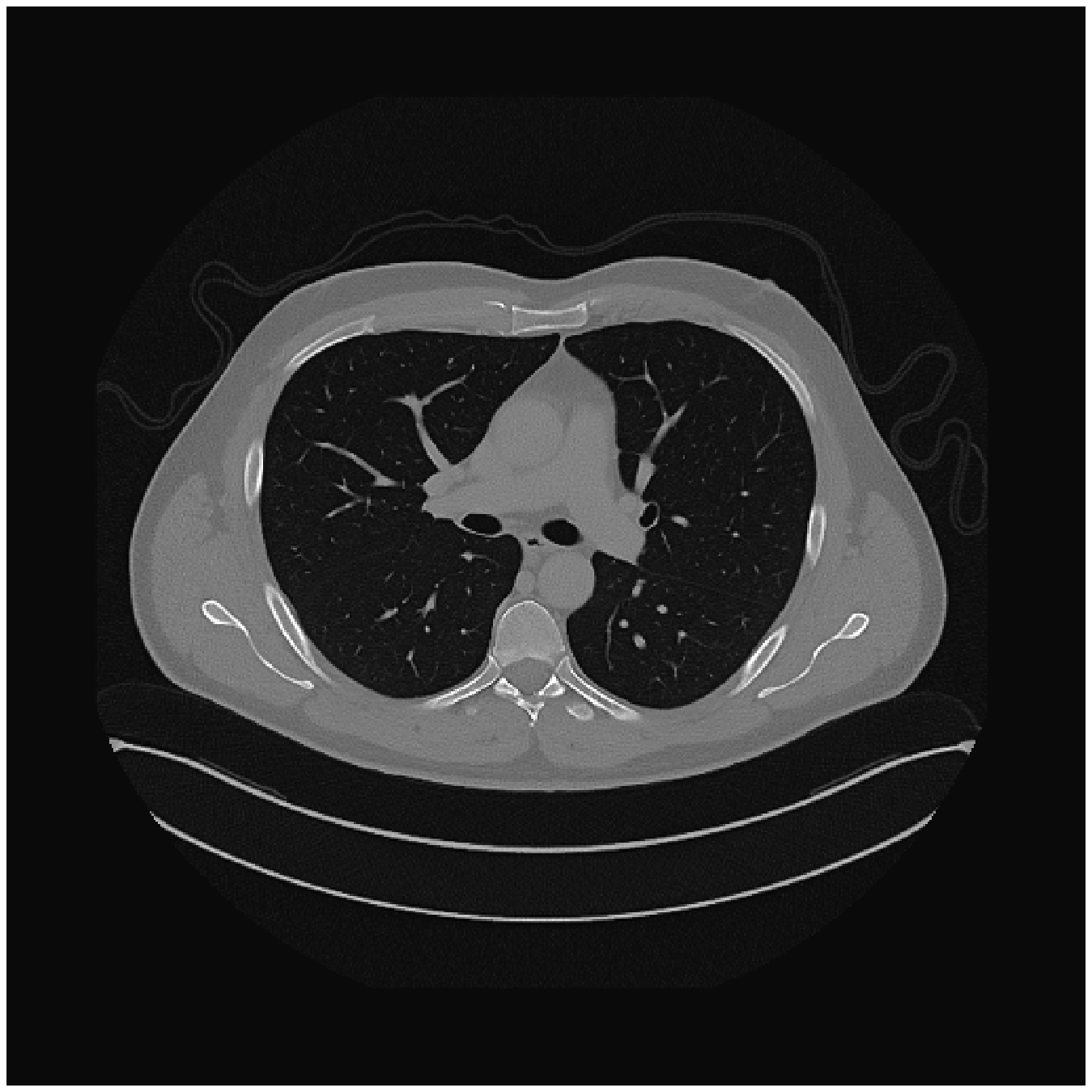}
\tiny\caption{Chest CT reference\label{fig:dref}}

\end{subfigure}
\begin{subfigure}{.24\textwidth}
\centering
  \includegraphics[trim={3.5cm 3.5cm 3.5cm 3.5cm},clip,width=4cm]{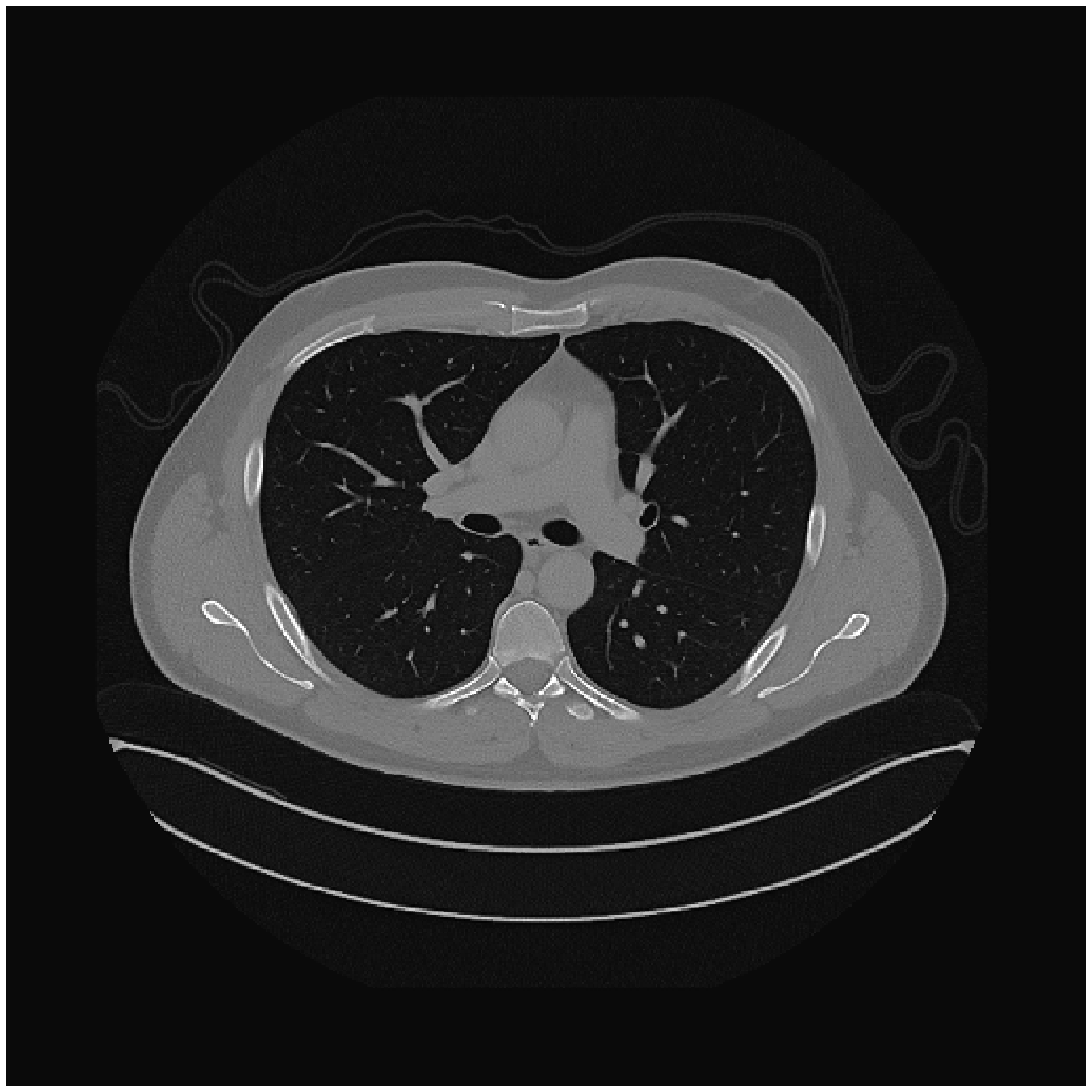}
\caption{QR reconstruction\label{fig:drec}}
\end{subfigure}
\vspace{-3pt}
\caption{CT images}
\label{fig:reconstructions}
%\vspace{-12pt}
\end{figure}

Table~\ref{tab:calidad} shows the quality metrics results, 
as an average of the quality of every slice of the phantom. The number of slices for every resolution in these tests is 32 for the $64^2$ pixels resolution, 128 for the $256^2$ and so on. The SSIM metric is equal to 1 for every image resolution, which indicates we are not losing any internal structure of the images. The PSNR is high for every case, with results always above 200, although it is higher for the smaller resolutions. In other works as \cite{CHILLARON20171195} where we worked with iterative methods, we considered reconstructions with a PSNR of around 60 to be high-quality.
Figures \ref{fig:reconstructions}.\subref{fig:ref} and \ref{fig:reconstructions}.\subref{fig:rec} show 
the central slice of the phantom and 
our reconstruction for a resolution with $512 \times 512$ pixels, the higher resolution we have reconstructed.
As can be observed, the images are identical.

A randomly chosen collection of real CT images from the dataset DeepLesion
\cite{deeplesion} was also tested. 
The selected images, which had $512 \times 512$ pixels, 
were projected with Joseph's method and used as reference. 
With these images from the dataset, 
the average PSNR of the reconstructions for 2048 slices 
corresponding to different studies is 220, and the SSIM is 1.
Figures \ref{fig:reconstructions}.\subref{fig:dref} and 
\ref{fig:reconstructions}.\subref{fig:drec}
show that our method achieves really high-quality reconstructions,
even though these images are much more complex than the phantom.
%\vspace{-5pt}

 \begin{table}[]
 \begin{minipage}[b]{0.48\linewidth}
\setlength\extrarowheight{3pt}
\setlength\tabcolsep{2pt}
\vspace{-7pt}
\caption{\small{Evolution of the relative residual.}}\label{tab:residual}
\vspace{-5pt}
\centering
\begin{tabular}{|c|c|c|c|c|}
\cline{2-5}
\multicolumn{1}{c}{} & \multicolumn{4}{|c|}{Resolution}  \\
\multicolumn{1}{c}{} & \multicolumn{4}{|c|}{}  \\
\cline{2-5}
\cline{2-5}
\multicolumn{1}{c|}{} & $64^2$ & $256^2$ & $384^2$ & $512^2$\\
\cline{2-5}
\hline
Residual & 
  $2.09 \cdot 10^{-13}$ & 
  $1.50 \cdot 10^{-12}$ & 
  $2.69 \cdot 10^{-12}$ & 
  $6.42 \cdot 10^{-12}$\\
%%%% \cline{2-5}
% \hline
% SSIM & 1 & 1 & 1 & 1 \\
\hline
\end{tabular}
%\vspace{-5pt}
\end{minipage}
%\end{table}
%\begin{table}[]
\begin{minipage}[b]{0.48\linewidth}
\setlength\extrarowheight{3pt}
\vspace{-7pt}
\caption{\small{Average Reconstruction Image Quality}}\label{tab:calidad}
\vspace{-5pt}
\centering
%\vspace{5pt}
\begin{tabular}{|c|c|c|c|c|}
\cline{2-5}
\multicolumn{1}{c}{} & \multicolumn{4}{|c|}{Resolution}  \\
\cline{2-5}
\cline{2-5}
\multicolumn{1}{c|}{} & $64^2$ & $256^2$ & $384^2$ & $512^2$\\
\cline{2-5}
\hline
PSNR & 258 & 228 & 220 & 204\\
%%%% \cline{2-5}
\hline
SSIM & 1 & 1 & 1 & 1 \\
\hline
\end{tabular}
%\vspace{-10pt}
\end{minipage}
\end{table}
%\vspace{-5pt}
\FloatBarrier
% -----------------------------------------------------------------------------
\subsection{Performance study}
\label{subsec:performances}
% -----------------------------------------------------------------------------
\vspace{-2pt}
The computer used in the performance experiments
featured one Intel i7-7800X\circledR\ CPU (6 physical cores)
and 128 GiB of RAM in total.
The clock frequency of the processor was 3.50 GHz,
and the so-called {\em Max Turbo Frequency} was 4.00 GHz.
In addition to one small SSD
for storing the operating system and programming tools,
the computer had two disks that were employed in the experiments,
both with a capacity of 2 TB:
One Hard Disk Drive (HDD) and 
one Solid-State Drive (SSD) with an M.2 connector.
The HDD (spinning disk) was a Toshiba DT01ACA200 (Firmware MX4OABB0).
The SSD was a Samsung SSD 970 EVO 2TB (Firmware 1B2QEXE7).
According to the Linux operating system \texttt{hdparm} tool,
the read speed of the first one was about 191.43 MB/s,
whereas the read speed of the second one was about was 2427.50 MB/s.
This is an upper-middle desktop personal computer and 
its current price is only about a few thousand dollars.
Its OS was GNU/Linux (kernel version 3.10.0-862.14.4.el7.x86\_64).
GCC compiler (version 4.8.5 20150623) was used.
Intel(R) Math Kernel Library (MKL)
Version 2018.0.2 Product Build 20180127 for Intel(R) 64 architecture
was employed for solving some advanced linear algebra problems.
Our new implementations were coded with 
the \texttt{libflame} (Release 11104) high-performance library,
which employed Intel's MKL for performing 
the small- and medium-sized basic linear algebra computations.

Because of the variability of the experimental running time on some computers, 
when solving linear systems 
three experiments were ran, and the average values were reported.
Nevertheless, we must say that 
the three obtained times were similar on the assessed architecture.
All the experiments reported show only 
the time required by the computation $X = R^{-1} (Q^T B)$,
since the QR factorization can be computed only once and 
then employed for many different images.

Unless explicitly stated otherwise,
all the experiments employed six threads (and therefore six cores)
for computation since the computer had six cores,
the only exception being the 
codes with overlapping of computation and I/O.
In this case, five threads (and five cores) were employed for computation
and one thread (one core) was employed for disk I/O tasks.

We have assessed four configurations, 
which are obtained as the combinations of 
two OOC AB methods (non-overlapping or basic OOC AB, and 
overlapping OOC AB) and
two types of disks (HDD and SSD).
The assessed four configurations were the following:

\begin{itemize}

\item
{\sc B-OOC + HDD}:
The basic (or non-overlapping) Out-Of-Core Algorithm-by-Blocks 
for solving the linear system was employed
on the HDD described above.
This is also called the initial configuration.

\item
{\sc O-OOC + HDD}:
The Out-Of-Core Algorithm-by-Blocks with overlapping of computation and I/O 
for solving the linear system was employed
on the HDD described above.

\item
{\sc B-OOC + SSD}:
The basic (or non-overlapping) Out-Of-Core Algorithm-by-Blocks 
for solving the linear system was employed
on the SSD described above.

\item
{\sc O-OOC + SSD}:
The Out-Of-Core Algorithm-by-Blocks with overlapping of computation and I/O 
for solving the linear system was employed
on the SSD described above.

\end{itemize}

In all our implementations we employed a block size 10240 
for the OOC computations
(the number of rows and columns of every square block, 
kept in a different file),
since this size usually renders good results 
on all the assessed code variants~\cite{ooc:qr,OOC:TOMS,ooc:supercomputing}.
In our codes,
the block size employed inside every task to process the blocks 
once they are stored in RAM was 128,
since this size usually renders good performances 
when processing matrices of size 10240.
In the rest of the codes not developed by us
(matrix-matrix products, etc.), 
the block size was determined by the library that performed that task 
(usually Intel's MKL).

\begin{figure}[t!]
%\tfvspace
\vspace{-20pt}
\centering
\includegraphics[width=0.48\textwidth]{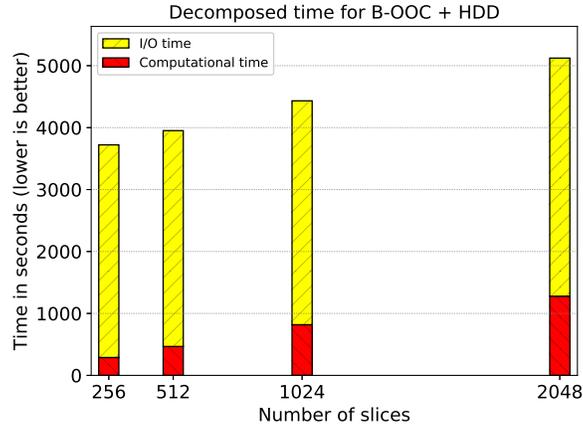} \\
\bfvspace
\vspace{10pt}
\caption{\small{Overall times and decomposed times of the initial configuration
(B-OOC + HDD)
for solving a linear system with A of dimension $266,500 \times 262,144$,
and B of dimension $266,500 \times k$, where $k$ is the number of slices.}}
\label{fig:decomposed_time_var4t_scratch}
\vspace{-10pt}
\end{figure}

Figure~\ref{fig:decomposed_time_var4t_scratch} 
shows the overall times and the decomposed times of the initial configuration
(B-OOC + HDD, that is, 
the basic or non-overlapping OOC Algorithm-by-Blocks on the HDD)
for solving a linear system with A of dimension $266,500 \times 262,144$,
and B of dimension $266,500 \times k$, where $k$ is the number of slices.
The aim of this plot was to assess if the process was feasible, and
to determine the main bottleneck of the application.
For the system with 2048 slices, 2.50 seconds per slice were needed;
for the system with 256 slices, 14.54 seconds per slice were needed.
These times showed that the process was feasible, 
but the times were a bit high in some cases and very high in other cases.
Moreover, the decomposition of the time showed that
I/O times were very high, but they did not grow too much 
as the number of slices increased.
Therefore, the main bottleneck of this problem was the I/O time,
instead of the computational time.
Then, adding more cores or several GPUs to the hardware configuration
was not going to help in this case,
and the focus should instead be on a fast disk.
\begin{figure}[b!]
\vspace{-11pt}
\begin{center}
%\onecolumn
%\begin{tabular}{c}
\includegraphics[width=0.48\textwidth]{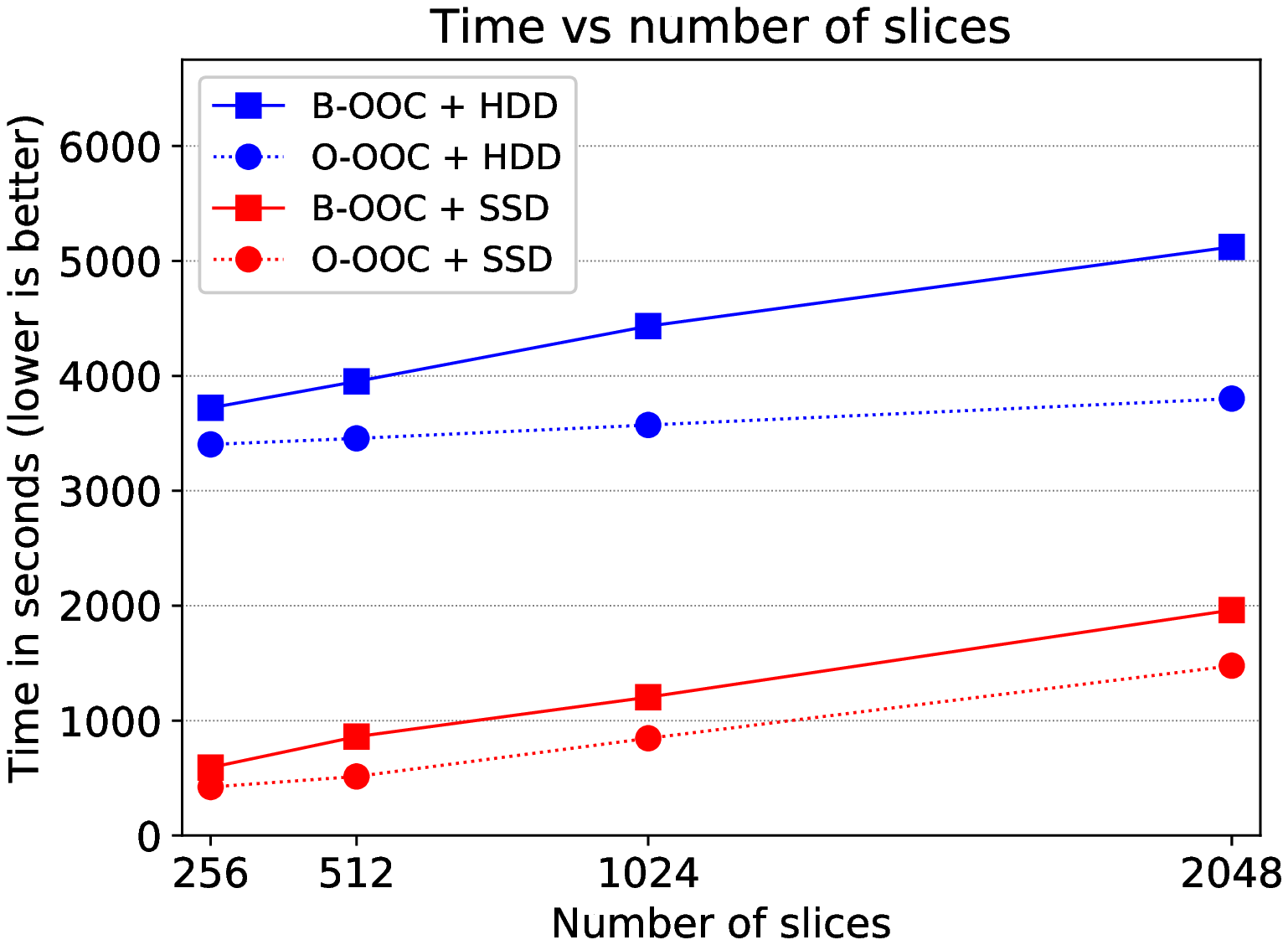}  \hfill \includegraphics[width=0.48\textwidth]{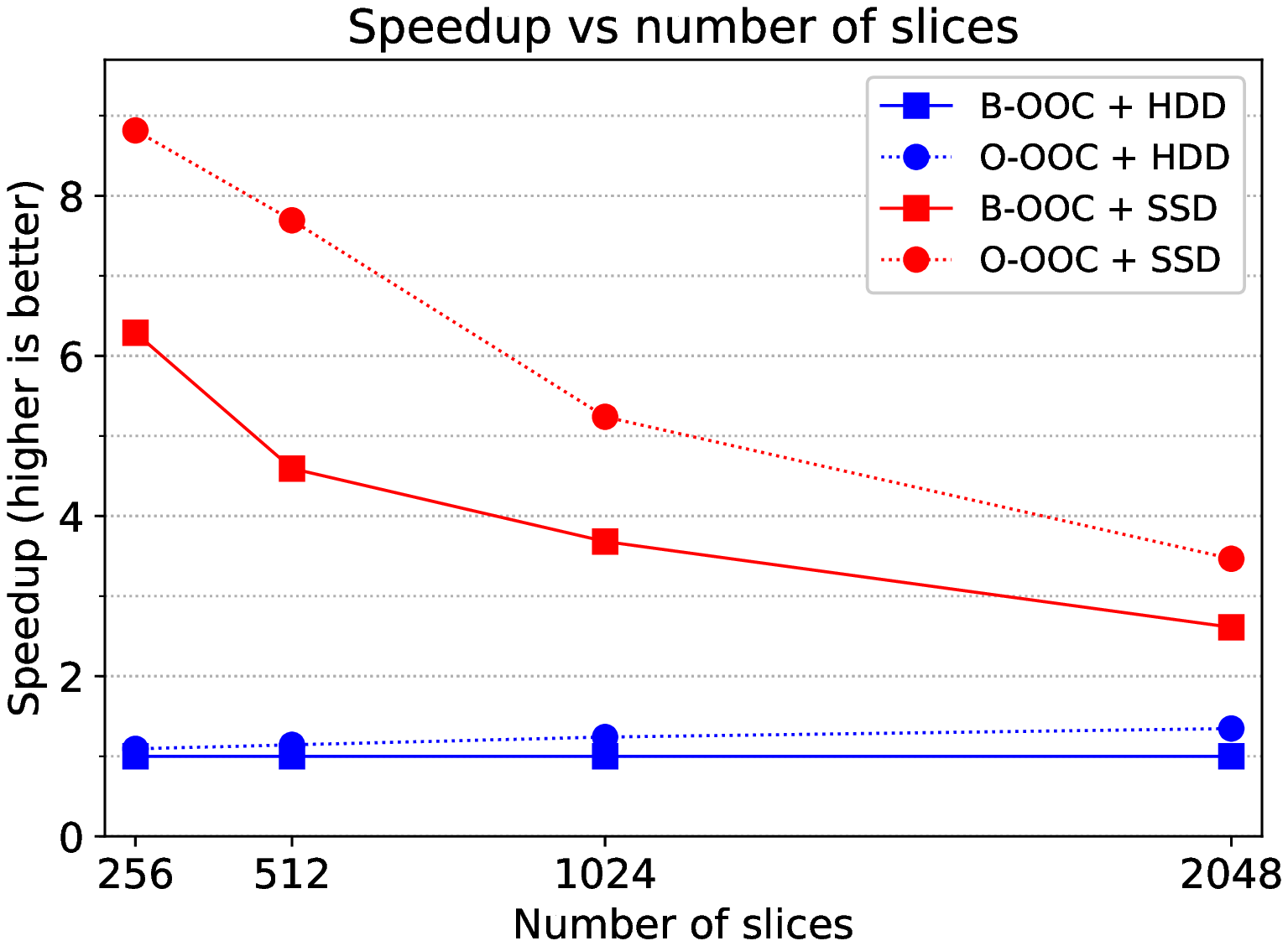} \\
%\end{tabular}
\end{center}
%\bfvspace
\vspace{-16pt}
\caption{\small{Time and speedups for the four configurations.}}
\label{fig:time_and_speedup}
 \vspace{-15pt}
\end{figure}
%\twocolumn
%\FloatBarrier
%

Figure~\ref{fig:time_and_speedup}
compares the performances of the four configurations
described above:
Basic OOC AB on HDD,
Overlapping OOC AB on HDD,
Basic OOC AB on SSD, and
Overlapping OOC AB on SSD.
The top subplot shows the times in seconds (lower is better),
whereas the bottom subplot shows the speedup (higher is better)
with respect to the initial configuration (basic OOC AB on HDD).
The speedup is computed as 
the quotient of the time obtained by the reference configuration
and the time obtained by the new configuration.
Thus, this concept means how many times the new configuration
is as fast as the reference configuration.
Hence, the higher the speedups, the better the performances are.
As the reference configuration is the initial one, 
in the bottom subplot the initial configuration will be shown as ones.
As can be seen, the SSD greatly reduced the overall times
and increased the speed by more than 6 times 
for the smallest case (256 slices) 
with respect to the initial configuration.
The overlapping of computation and I/O further increased 
the speed up to nearly 9 times for the smallest case (256 slices).
When the number of slices was high,
the improvements were not so great but still very noticeable.

\begin{figure}[t!]
\vspace{-5pt}
\begin{minipage}[c]{0.49\linewidth}
%\vspace{-10pt}
\begin{center}
\includegraphics[width=1\textwidth]{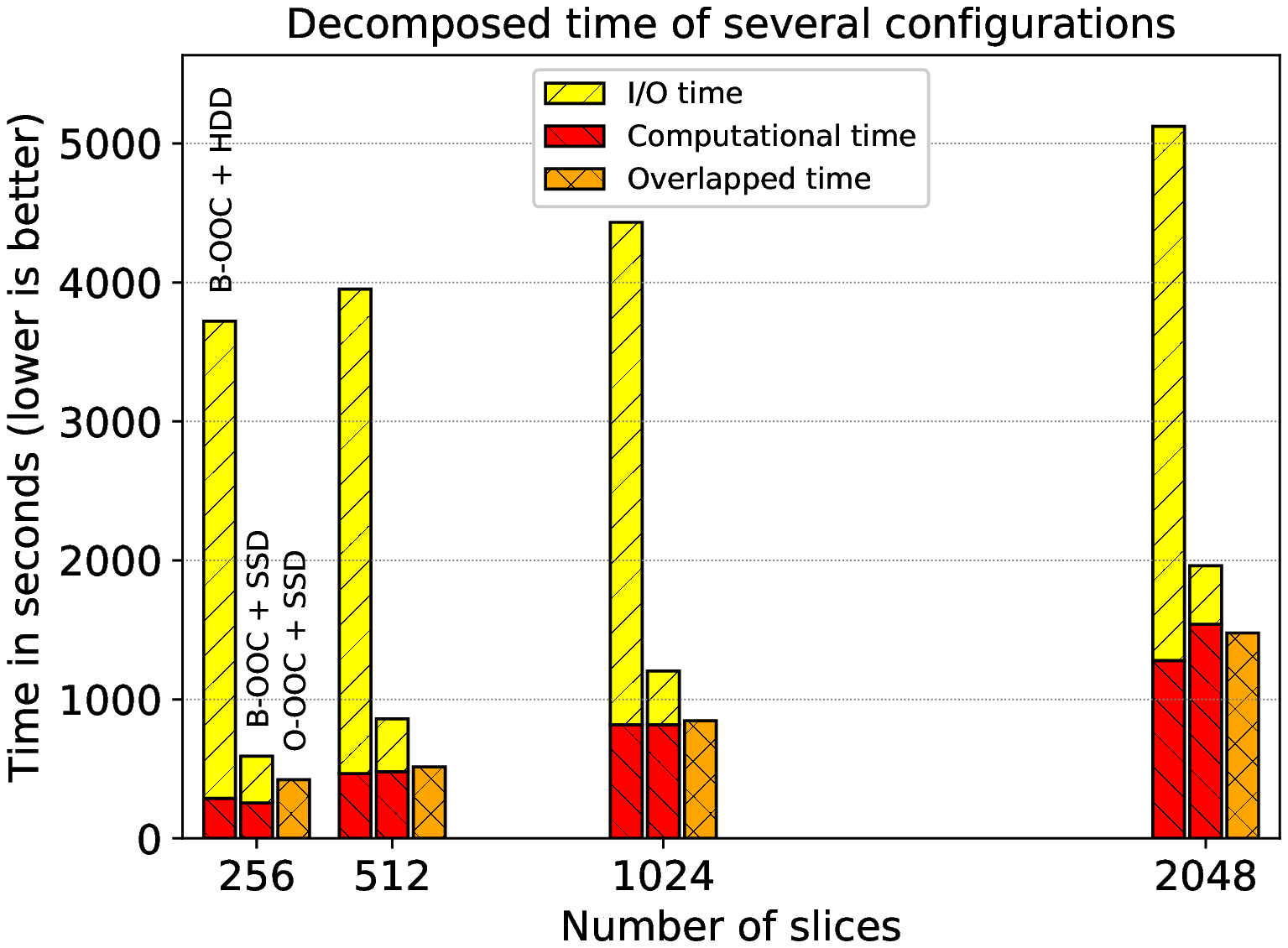} \\
\end{center}
\vspace{-15pt}
\caption{\small{Overall times and decomposed times of three configurations
for solving a linear system with A of dimension $266,500 \times 262,144$,
and B of dimension $266,500 \times k$, where $k$ is the number of slices.}}
\label{fig:combined_decomposed_time}
%\vspace{-12pt}
\end{minipage}
\hfill
\begin{minipage}[c]{0.49\linewidth}
\vspace{-30pt}
\begin{center}
\includegraphics[width=1\textwidth]{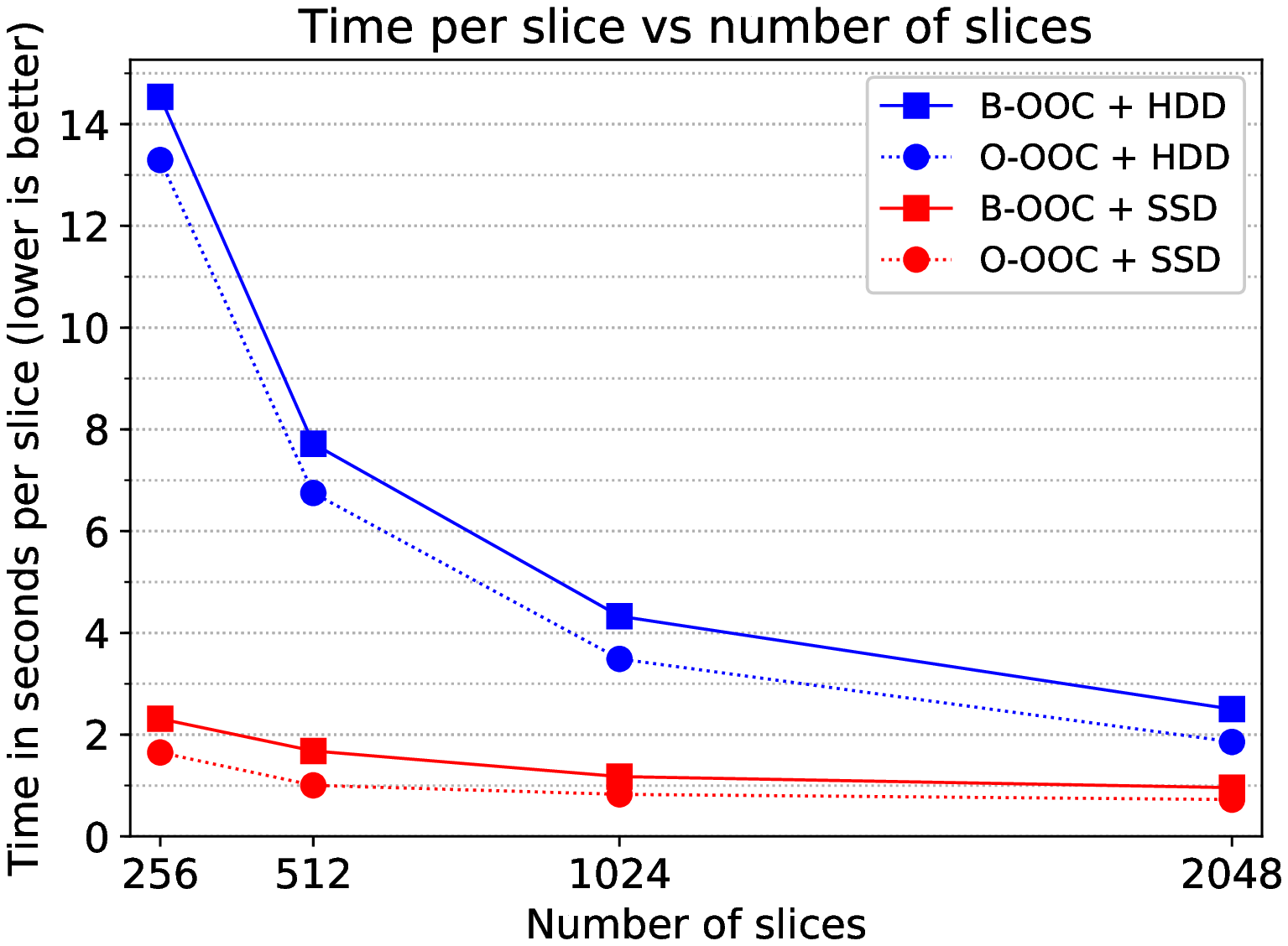} \\
\end{center}
\vspace{-15pt}
\caption{\small{Time in seconds per slice for the four configurations.}}
\label{fig:time_per_slice}
%\vspace{-12pt}
\end{minipage}
\vspace{-5pt}
\end{figure}

Figure~\ref{fig:combined_decomposed_time}
shows the overall times and the decomposed times
for solving a linear system with A of dimension $266,500 \times 262,144$,
and B of dimension $266,500 \times k$, where $k$ is the number of slices,
on three configurations: B-OOC + HDD, B-OOC + SSD, and O-OOC + SSD.
The left bar for each number of slices 
shows the overall times and the decomposed times
of the initial configuration (B-OOC + HDD).
As can be seen, 
its main drawback is the high I/O cost because of using a HDD. 
The center bar for each number of slices 
shows the overall times and the decomposed times of a configuration similar to the previous one with an SSD (B-OOC + SSD).
As can be seen, 
the high I/O cost has been greatly reduced.
The right bar for each number of slices 
shows the overall times
of the best configuration (O-OOC + SSD).
As this configuration overlaps computation and I/O, 
the time cannot be decomposed.
As can be seen, 
in most cases the I/O cost (the fast SSD) of the previous configuration is 
completely removed.
%\FloatBarrier

Figure~\ref{fig:time_per_slice}
shows the time in seconds required to compute one slice.
As it shows, this time was not constant, and
it depended somewhat on the number of slices:
the more slices to compute, the lower the time per slice.
Just consider that, regardless of the number of slices 
(even for just one slice),
the whole factorized matrix $A$ must be read from disk.
Thus, this large cost becomes diluted as more slices are being computed.
In the initial configuration (basic OOC AB and HDD) 
the time per slice greatly depended on the number of slices.
In the best configuration (overlapping OOC AB on SSD)
the time per slice is not so dependent on the number of slices.

Table~\ref{tab:time_per_slice}
shows the time in seconds required to compute one slice
for both the initial configuration and 
the most performant configuration.
%\FloatBarrier

As can be seen, the range of the initial configuration 
is very wide (from 2.50 to 14.54 seconds),
whereas the range of the most performant configuration
is much narrower (from 0.72 to 1.65 seconds).

The weights matrix for the highest resolution in our experimental study
required a storage of about 560 GB 
($265,500 \times 262,144$ double precision elements).
Besides the weights matrix,
additional space
(patient's data, final image, temporary data,
application code, operating system, disk buffers and cache, etc.)
makes the total size required by this problem even larger.
As was told, the computer had 128 GB of RAM.
However, only 32 GB were employed as a cache to store
blocks of the weights matrix,
leaving the rest for other purposes 
(operating system disk cache and buffers, etc.).
We assessed another computer with 48 GB of RAM, and results were similar
when using a similar number of cores,
but we did not report its results because it was a much more expensive server.
Hence, to obtain good performances,
a smaller main memory could be used (thus reducing the total price),
but a fast SSD is a strong requirement.

\begin{table}[b!]
%\tfvspace
\vspace{-15pt}
%\vspace{20pt}
\begin{center}
\caption{\small{Time in seconds per slice versus number of slices.}}
\label{tab:time_per_slice}
\vspace{-2pt}
\begin{tabular}{|l|c|c|c|c|}
\cline{2-5}
  \multicolumn{1}{c}{} &
  \multicolumn{4}{|c|}{Number of slices} \\ \cline{1-5}
  \multicolumn{1}{|c|}{Method} & 256   & 512  & 1024 & 2048 \\ \hline
  B-OOC + HDD                  & 14.54 & 7.72 & 4.33 & 2.50 \\
  O-OOC + SDD                  &  1.65 & 1.00 & 0.83 & 0.72 \\ \hline
\end{tabular}
\end{center}
%%%% \bfvspace
\vspace{-10pt}
\end{table}

%\vfill
\FloatBarrier

%% file: 4_Conclusions.tex
% =============================================================================
\section{Conclusions}
\label{sec:conclusions}
% =============================================================================

In this paper, we present a direct algebraic method based on the QR
factorization for reconstructing CT images efficiently on affordable computers.
As we have shown, 
this method is numerically stable even for high resolutions 
provided the weights matrices have full rank.
For this reason, 
our method employs more X-ray projections than the iterative methods,
but fewer than the analytical methods. 
Besides, the use of few projections can affect the result of the diagnosis 
since some pathologies could be hidden or simulated. 

With our proposed method, 
which uses a number of projections 
that guarantee the full rank of the weights matrix, 
high-quality images are obtained without requiring an a-priori knowledge or 
interaction with the patient.
This method guarantees the non-creation of artifacts 
except those produced by problems on detectors, dispersion, movement, 
intensity of the source, etc., 
which can be corrected by filtering and segmentation techniques.
In addition, our reconstructions achieve remarkable quality 
even for complex real CT images.
It is worth to mention we have not considered or removed the possible noise
on the projections, which is left as a future work.

We have shown that an efficient reconstruction of CT images
can be achieved using Out-Of-Core and Algorithm-By-Blocks techniques.
By using our techniques, 
affordable computers with a price of about one order of magnitude lower
can be successfully employed,
because a large main memory (which is quite expensive) 
is not required, just a fast hard drive.
For this reason, the equipment needed to reconstruct the images 
is affordable and thus more accessible to the public.
The type of hard drive can improve our reconstruction times drastically.
When using a HDD, the performance is dominated by the I/O time, 
whereas when using an SSD the I/O time is greatly reduced and 
the performance is dominated by the computational time again.
Furthermore, the method that overlaps computation and I/O
can further reduce the reconstructing time, 
thus making our method more competitive. We could perform an standard CT study with resolution $512^2$ and 256 slices in about 4 minutes.
We have also shown that the cost per slice is lower as the number of
simultaneous slices to reconstruct is higher, 
which would be beneficial for full-body CT scans.

Considering all of the above,
we conclude that computing high-quality reconstructions 
with direct algebraic methods
on affordable equipment can be achieved.
Since our method is stable we could increase the resolution of the images
provided we had enough storage space, and still get valid results.
\vspace{30pt}